\documentstyle[preprint,epsbox]{jpsj}

\def\GA{\raise2.5pt\hbox{$>$}\kern-8pt\lower2.5pt\hbox{$\sim$}}
\def\LA{\raise2.5pt\hbox{$<$}\kern-8pt\lower2.5pt\hbox{$\sim$}}
\def\runtitle{Correlated Insulating State}
\def\runauthor{Takuya OKABE}

\title
{
Generalization of  Gutzwiller Approximation
}

\author
{ 
Takuya {\sc Okabe}
\footnote{E-mail: okabe@ton.scphys.kyoto-u.ac.jp}
}
\inst
{
Department of Physics, Kyoto University, Kyoto 606-01
}

\recdate
{
January 20, 1997
}

\abst
{We derive expressions required 
in generalizing the Gutzwiller approximation
to models comprising arbitrarily degenerate localized orbitals.
}

\kword
{
Gutzwiller approximation,
degenerate-band model
}

\begin{document}
\sloppy
\maketitle

Previously, we generalized
the Gutzwiller approximation 
to degenerate-band models, and
investigated itinerant ferromagnetism in 
the $3d$ transition metal systems.~\cite{rf:Okabef}\
There, we presented the resulting formula
but omitted its derivation because it was quite
complicated.~\cite{rf:comment1}\
However, once we know the physical interpretation 
of the result as explained in ref.~\citen{rf:Okabef},
it is natural to expect that 
the formula may be derived rather straightforwardly
in the way that it reflects the simple interpretation.
In fact, as we found this is actually the case,
we report it in the following.
At the end, we show that the Brinkman-Rice 
transition occurs only when the carrier density 
equals an integer.

We investigate the Hamiltonian comprising arbitrary numbers
of localized orbitals $l$;
\begin{equation}
H=T+V=\sum_{i,j,l}t^{l}_{ji}c^\dagger_{jl}c_{il}+
\sum_i
\sum_{
{\scriptstyle p
}\atop
{\scriptstyle \{l_{1}, \cdots , l_{p}\}}}
C^{(p)}(l_{1}, \cdots , l_{p})\hat{\nu}_i^{(p)}(l_{1}, \cdots , l_{p}).
\label{eq:V}
\end{equation}
For simplicity,
the hopping part $T$ is assumed to be diagonal with respect to $l$.
In the potential part $V$,
indices $l_m$  designate
orbital and spin of localized states.
Operators $\hat{\nu}_i^{(p)}(l_{1}, \cdots , l_{p})$
have eigenvalue 1 if only $p$ orbitals $l_{1}, \cdots , l_{p}$
are occupied (the others unoccupied) at site $i$.
These operators correspond to 
$\hat{d}_i\equiv \hat{n}_{i\uparrow}\hat{n}_{i\downarrow}$
of the single band Hubbard model.
The sum in $V$ is taken 
over a set $\{l_{1}, \cdots , l_{p}\}$,
not to count their combination redundantly.
The Pauli principle requires
$l_m \neq l_n$ for $m\ne n$ 
for the orbitals in the braces $\{l_{1}, \cdots , l_{p}\}$.
$C^{(p)}(l_{1}, \cdots , l_{p})$ 
is  interaction energy of an eigenstate
of $\hat{\nu}_i^{(p)}(l_{1}, \cdots , l_{p})$.
Generalizing  the Gutzwiller approximation,
we can estimate the expectation value of $H$ as
\begin{equation}
\frac{1}{L}\langle\Psi|H|\Psi\rangle=
\sum_{l} q(l)\tilde{\varepsilon}_{l}
+\sum_{{\scriptstyle p
}\atop{\scriptstyle \{l_{1}, \cdots ,
l_{p}\}}}
C^{(p)}(l_{1}, \cdots , l_{p})\nu^{(p)}(l_{1}, \cdots , l_{p}),
\label{Epersite}
\end{equation}
where $\tilde{\varepsilon}_{l}$ is defined by
\begin{equation}
\tilde{\varepsilon}_{l}\equiv \frac{1}{L}
\langle\Psi_0|\sum_{i,j}
t^{l}_{ji}c^\dagger_{jl}c_{il}|\Psi_0\rangle,
\label{AvKin}
\end{equation}
and represents the average kinetic energy 
of the band $l$ for the uncorrelated state $|\Psi_0\rangle$,
for which we assume $\langle\Psi_0|\Psi_0\rangle=1$.
In eqs.~(\ref{Epersite}) and (\ref{AvKin}),
$L$ is the total number of lattice sites.
Parameters $\nu^{(p)}(l_{1}, \cdots , l_{p})$
are the expectation value
of $\hat{\nu}_i^{(p)}(l_{1}, \cdots , l_{p})$,
and take non-negative values.
In the uncorrelated case, where $C^{(p)}=0$ for any $p$,
these are given by
\begin{equation}
\nu^{(p)}(l_{1}, \cdots , l_{p})=
\nu^{(p)}_0(l_{1}, \cdots , l_{p})\equiv
\prod_{i=\{1,\cdots,p\}} n(l_{i})
\prod_{i\ne \{1,\cdots,p\}} \left(1-n(l_{i})\right),
\label{nu0}
\end{equation}
where
\begin{equation}
n(l)\equiv \frac{1}{L}
\langle\Psi_0|\sum_{i}c^\dagger_{il}c_{il}|\Psi_0\rangle.
\end{equation}

Most difficult and laborious part in applying 
the Gutzwiller approximation is
the calculation
of the band-width reduction factor $q(l)$
in eq.~(\ref{Epersite})
as a function of $\nu^{(p)}(l_{1}, \cdots , l_{p})$.
If this is achieved, $\nu^{(p)}(l_{1}, \cdots , l_{p})$
are determined so as to minimize $\langle\Psi|H|\Psi\rangle$.
In the variational calculation, 
these parameters $\nu^{(p)}$ cannot take arbitrary values since they 
must satisfy two relations,~\cite{rf:Okabef}\
i.e.,
`conservation of number',
\begin{equation}
n(l_{1})=
\nu^{(1)}(l_{1})+\sum_{p(\geq2)}
\sum_{\{l_{2},\cdots,l_{p}\}(\ne l_1)}
\nu^{(p)}(l_{1},l_{2},\cdots,l_{p}),
\label{n-conserv}
\end{equation}
and 
`conservation of  probability',
\begin{equation}
\sum_{p}  \sum_{\{l_{1}, \cdots, l_{p}\}}
 \nu^{(p)}(l_{1}, \cdots ,l_{p})=1.
\label{prob-conserv}
\end{equation}
Therefore for fixed $n(l_i)$ we must regard  
only $\nu^{(p)}$ with $p\ge 2$ as
variational parameters.

\begin{figure}
\centerline{\epsfile{file=fig1,width=6.6cm}}
\caption{Examples of the hopping process.
A carrier in the orbital $l_1$ at site $i$ moves to 
the orbital $l_{1'}$ at site $j$.
}
\label{fig:process}
\end{figure}
A general result for $q(l)$
was obtained
for the first time in ref.~\citen{rf:Okabef},
\begin{equation}
q(l_{1})=\frac{1}{n(l_{1})\left(1-n(l_{1})\right)}
\left( \sum_{p \ge 1}\sum_{\{l_{2},\cdots,l_{p}\}(\neq l_1)}
\sqrt{\nu ^{(p)}(l_{1}, \cdots , l_{p})}
\sqrt{\nu ^{(p-1)}(l_{2}, \cdots , l_{p})} \right)^{2},
\label{Gqfactor} 
\end{equation} 
and interpreted physically as shown in Fig.~\ref{fig:process}.\
Terms in the round bracket squared are interpreted as 
a sum ($\sum_p$) of products 
$\bigl(\sqrt{\nu ^{(p)}}\sqrt{\nu ^{(p-1)}}\bigr)$
of probability amplitudes $\bigl(\sqrt{\nu ^{(p)}}\bigr)$,
which are involved in the hopping process under consideration.
The square is due to contributions from two sites $i$ and $j$. 
One can make sure that 
$\nu^{(p)}_0(l_{1}, \cdots , l_{p})$ defined in
eq.~(\ref{nu0}) meet eqs.~(\ref{n-conserv}) and (\ref{prob-conserv}),
and thus verify that $q(l)=1$ for $C^{(p)}=0$
using eq.~(\ref{prob-conserv}), as expected.
To derive eq.~(\ref{Gqfactor}),
we must derive relations between $\nu^{(p)}$
and the Gutzwiller projection factors $\eta^{(p)}$ ($\le 1$),
which are defined in 
the generalized Gutzwiller wavefunction,
\begin{equation}
|\Psi\rangle=
\prod_{
{\scriptstyle i} \atop {\scriptstyle p, \{l_1,\cdots,l_p\}}}
\left[ 1-\left(1-\eta^{(p)}(l_1,\cdots,l_p)\right)\hat{\nu_i}^{(p)}
(l_1,\cdots,l_p)\right]|\Psi_0\rangle,
\label{gGWF}
\end{equation}
where $\eta^{(1)}(l_i)\equiv \eta^{(0)}\equiv 1$.

In the Gutzwiller approximation,
the norm of the state (\ref{gGWF}) is estimated as
\begin{eqnarray}
\langle \Psi|\Psi\rangle=
\prod_{{\scriptstyle p} \atop {\scriptstyle \{l_1,\cdots,l_p\}}}
\sum_{ \bar{\nu}^{(p)}}\,
\left[\eta^{(p)}(l_1,\cdots,l_p)\right]
^{2\bar{\nu}^{(p)}(l_1,\cdots,l_p)}
N\left(L,\{N(l_i)\},\{\bar{\nu}^{(p)}\}\right)P\left(L,\{N(l_i)\}
\right),
\label{norm}
\end{eqnarray}
where 
\begin{eqnarray}
\bar{\nu}^{(p)}(l_1,\cdots,l_p)&\equiv &
L\,\nu^{(p)}(l_1,\cdots,l_p),\nonumber\\
N(l_i)&\equiv& L\,n(l_i),\nonumber \\
N\left(L,\{N(l_i)\},\{\bar{\nu}^{(p)}\}\right)
&=&\frac{L!}{\displaystyle{\prod_{{\scriptstyle p} \atop
{\scriptstyle \{l_1,\cdots,l_p\}}}
\strut \bar{\nu}^{(p)}(l_1,\cdots,l_p)\,!}},\nonumber \\
P\left(L,\{N(l_i)\} \right)&=&\prod_i
n(l_i)\strut ^{N(l_i)}\left(1-n(l_i)\right)^{L-N(l_i)}.
\end{eqnarray}
These are generalization of 
the results of Vollhardt~\cite{rf:Vollhardt}
for the single band model.

Then, we shall approximate
the sum over $\bar{\nu}^{(p)}$ in eq.~(\ref{norm})
by a single term which gives 
the largest contribution in the thermodynamic limit $L\rightarrow\infty$;
$\bar{\nu}^{(p)}$ ($p\ge 2$) is determined by 
\begin{equation}
\frac{{\rm d}}{{\rm d}\bar{\nu}^{(p)}(l_1,\cdots,l_p)}
\log \langle \Psi|\Psi\rangle
=0,
\label{dlogdnu=0}
\end{equation}
from which we obtain a relation between 
$\eta^{(p)}(l_1,\cdots,l_p)$ and $\nu^{(p)}(l_1,\cdots,l_p)$,
\begin{equation}
\left[\eta^{(p)}(l_1,\cdots,l_p)\right]^2
=\frac{\strut(\nu^{(0)})^{p-1}\nu^{(p)}(l_1,\cdots,l_p)}{
\strut\nu^{(1)}(l_1)\cdots\nu^{(1)}(l_p)}.
\label{etanu}
\end{equation}
To derive eq.~(\ref{etanu}) from eqs.~(\ref{norm}) and (\ref{dlogdnu=0}),
one may note 
\begin{eqnarray}
%
\frac{{\rm d}\bar{\nu}^{(1)}(l_i)}{{\rm d}\bar{\nu}^{(p)}(l_1,\cdots,l_p)}
&=&-1, \qquad (i=1,\cdots,p)\nonumber\\
\frac{{\rm d}\bar{\nu}^{(0)}}{{\rm d}\bar{\nu}^{(p)}(l_1,\cdots,l_p)}
&=&p-1,
\end{eqnarray}
owing to eqs.~(\ref{n-conserv})
and (\ref{prob-conserv}),
and 
\begin{equation}
\frac{{\rm d}}{{\rm d}N}\log N!=\log N. \qquad (N\rightarrow\infty)
\end{equation}


As a next step,
we note that 
spatial correlation of various configurations
is completely neglected in the Gutzwiller approximation.
Thus,
the factor $q$ can be separated into two independent parts,
each of which comes from contribution for the creation 
and annihilation operators;
\begin{equation}
\frac{\langle\Psi|c^\dagger_{jl_{1'}}c_{il_1}|\Psi\rangle}
{\langle\Psi|\Psi\rangle}
\rightarrow
r_{c^\dagger_{l_{1'}}}
r_{c^{\rule{0cm}{1ex}}_{l_1}}
\frac{\langle\Psi_0|c^\dagger_{jl_{1'}}c_{il_1}|\Psi_0\rangle}{
\langle\Psi_0|\Psi_0\rangle}.
\label{split}
\end{equation}
To calculate $r_{c^{\rule{0cm}{1ex}}_{l_1}}$
for the process shown in Fig.~\ref{fig:process},
two factors appear
besides  the Gutzwiller parameters
$\eta^{(p)}\left(\left\{ l_1,\cdots,l_{p}\right\}\right)
\eta^{(p-1)}\left(\{l_2,\cdots,l_{p}\}\right)$.
These are the following ratios,  i.e.
\begin{eqnarray}
\frac{N\left(L-1,N(l_1)-1,N(l_{2})-1,\cdots,N(l_p)-1;\cdots\right)}{
N\left(L,N(l_1),N(l_{2}),\cdots,N(l_p);\cdots\right)}
=\frac{\nu^{(1)}(l_1)\nu^{(1)}(l_{2})\cdots\nu^{(1)}(l_p)}{
\strut (\nu^{(0)})^{p-1}},
\label{Nratio}
\end{eqnarray}
and
\begin{equation}
\frac{P\left(L-1,N(l_1)-1\right)}{P\left(L,N(l_1)\right)}
=\frac{1}{n(l_1)},
\label{Pratio}\end{equation}
where dots after the semicolon in 
the denominator and numerator of the
left-hand side of eq.~(\ref{Nratio}) 
represent common parts.
We need these ratios since
the orbital $l_1$ at the site $i$,
where orbitals $l_2, \cdots, l_p$ are occupied,
should also be occupied before the hopping process.
Therefore the other $N(l_i)-1$ $(i=1,\cdots,p)$
carriers must be on the other $L-1$ sites.
%

Finally, we obtain the result for $r_{c_{l_1}}$,
\begin{eqnarray}
\label{annihilate}
r_{c_{l_1}}
&=&\frac{\nu^{(1)}(l_1)}{n(l_1)}
\biggl[
 1
+\nonumber \\
&& 
+\sum_{
  {\scriptstyle p\ge 2} \atop
  {\scriptstyle \{l_2,\cdots,l_{p}\} (\ne l_1)}
  }
\eta^{(p)}
\left(
 \left\{ l_1,\cdots,l_{p}
 \right\}
 \right)
\eta^{(p-1)}
\left(
 \{l_2,\cdots,l_{p}\}
 \right) 
\frac{\nu^{(1)}(l_2)\cdots\nu^{(1)}(l_{p})}{(\nu^{(0)})^{p-1}}
\biggr].
\end{eqnarray}

In the same fashion, in terms of
\begin{eqnarray}
\frac{N\left(L-1,N(l_{1'}),N(l_{2'})-1,\cdots,N(l_{p'})-1
;\cdots\right)}{N\left(L,N(l_{1'})
,\cdots,N(l_{p'});\cdots\right)}
=\frac{\nu^{(0)}\nu^{(1)}(l_{2'})\cdots\nu^{(1)}(l_{p'})}{\strut
(\nu^{(0)})^{p'-1}},
\end{eqnarray}
and
\begin{equation}
\frac{P\left(L-1,N(l_{1'})\right)}{P\left(L,N(l_{1'})\right)}
=\frac{1}{1-n(l_{1'})},
\end{equation}
one obtains
\begin{eqnarray}
\label{create}
r_{c^\dagger_{l_{1'}}}
&=&\frac{\nu^{(0)}}{1-n(l_{1'})}
\biggl[
 1+\nonumber\\
&&+\sum_{
  {\scriptstyle p'\ge 2} \atop
  {\scriptstyle \{l_{2'},\cdots,l_{p'}\} (\ne l_{1'})}
  }
\eta^{(p')}
\left(
 \left\{ l_{1'},\cdots,l_{p'}
 \right\}
 \right)
\eta^{(p'-1)}
\left(
 \{l_{2'},\cdots,l_{p'}\}
 \right)
\frac{\nu^{(1)}(l_{2'})\cdots\nu^{(1)}(l_{p'})}{(\nu^{(0)})^{p'-1}}
\biggr].
\end{eqnarray}
Substituting eq.~(\ref{etanu}) to 
eqs.~(\ref{annihilate}) and (\ref{create}) for
$l_1=l_{1'}=l$,
we finally conclude the result (\ref{Gqfactor})
for $q(l)=r_{c^\dagger_{l}}r_{c^{\rule{0cm}{1ex}}_{l}}$.

Generally in the case $l\ne l'$,
asymmetry arises between 
$r_{c^\dagger_{l'}}$ and $r_{c^{\rule{0cm}{1ex}}_{l}}$,
or, one obtains different factors $q_{l'l}$ and $q_{ll'}$ for
$\langle c^\dagger_{jl'}c_{il}\rangle$ 
and $\langle c^\dagger_{il}c_{jl'}\rangle$, respectively.
These should be the same because of the Hermite character
of the Hamiltonian.
Thus in this situation, one may instead use their average 
$q=(q_{l'l}+q_{ll'})/2$ for both of 
$\langle c^\dagger_{jl'}c_{il}\rangle$ 
and $\langle c^\dagger_{il}c_{jl'}\rangle$.
In the same manner,
the $q$ factors for 
$\langle c^\dagger_{jl'}c^\dagger_{il}\rangle$
and $\langle c_{il}c_{jl'}\rangle$
take different forms.
In this case, however, one can use their geometrical mean
since these factors always appear pairwise as a product
in physical terms.

Notwithstanding general treatment, 
compared with the method of Vollhardt,~\cite{rf:Vollhardt}\
our derivation presented above 
is not only simple but transparent with respect to 
the physical meaning of respective terms
contributing to the $q$ factor.
This is due to observation made at (\ref{split});
we treated the two sites involved separately,
while usually these are treated altogether.
In particular,
the square root $\sqrt{\nu^{(p)}}$
appears in eq.~(\ref{Gqfactor})
in place of $\eta^{(p)}$, eq.~(\ref{etanu}),
while
squared $\eta^{(p)}$ there 
stems from the norm $\langle\Psi|\Psi\rangle$, eq.~(\ref{norm}).
Therefore our derivation validates our previous interpretation 
of $q(l)$ as a sum of products of probability amplitudes
$\sqrt{\nu^{(p)}}$.

In passing,
we end this article with a comment on 
the metal-insulator transition of the type noted first by 
Brinkman and Rice.~\cite{rf:BR}\
Summing eq.~(\ref{n-conserv}) over $l_1$,
we obtain 
\begin{eqnarray}
n&=&\sum_{l_1}n(l_1)=\sum_{l_1}\nu^{(1)}(l_1)+\sum_{l_1}
\sum_{p(\ge 2)}
\sum_{\{l_{2},\cdots,l_{p}\} (\ne l_1)}
\nu^{(p)}(l_{1},l_{2},\cdots,l_{p})\nonumber\\
&=&\sum_{l_1}\nu^{(1)}(l_1)
+\sum_{p(\ge 2)}
\sum_{\{l_{1},\cdots,l_{p}\}}
p\, \nu^{(p)}(l_{1},l_{2},\cdots,l_{p})\nonumber\\
&=&\sum_{p}
\sum_{\{l_{1},\cdots,l_{p}\}}
p\, \nu^{(p)}(l_{1},l_{2},\cdots,l_{p}).
\label{totaln}
\end{eqnarray}
If the on-site interaction is strong enough,
one may consider only configurations which have $p-1$, $p$
and $p+1$ carriers on a single site,
where an integer $p$ is selected so that $p-1< n< p+1$.
Then,
if we set probabilities of the other configurations as zero,
eqs.~(\ref{totaln}), (\ref{prob-conserv}) and 
(\ref{Gqfactor}) become
\begin{eqnarray}
n&=&\sum_{\{l_{1},\cdots,l_{p-1}\}}
(p-1)\, \nu^{(p-1)}(l_{1},l_{2},\cdots,l_{p-1})
+
\sum_{\{l_{1},\cdots,l_{p}\}}
p\, \nu^{(p)}(l_{1},l_{2},\cdots,l_{p}) \nonumber \\
&&+
\sum_{\{l_{1},\cdots,l_{p+1}\}}
(p+1)\, \nu^{(p+1)}(l_{1},l_{2},\cdots,l_{p+1}),
\label{nend}\\
1&=&\sum_{\{l_{1},\cdots,l_{p-1}\}}
\nu^{(p-1)}(l_{1},l_{2},\cdots,l_{p-1})
+
\sum_{\{l_{1},\cdots,l_{p}\}}
\nu^{(p)}(l_{1},l_{2},\cdots,l_{p}) \nonumber \\
&&+
\sum_{\{l_{1},\cdots,l_{p+1}\}}
\nu^{(p+1)}(l_{1},l_{2},\cdots,l_{p+1}),
\label{probend}
\end{eqnarray}
and 
\begin{eqnarray}
q(l_{1})&=&\frac{1}{n(l_{1})\left(1-n(l_{1})\right)}
\Biggl( 
\sum_{\{l_{2},\cdots,l_{p}\}(\neq l_1)}
\sqrt{\nu ^{(p)}(l_{1}, \cdots , l_{p})}
\sqrt{\nu ^{(p-1)}(l_{2}, \cdots , l_{p})} \nonumber\\
&&+
\sum_{\{l_{2},\cdots,l_{p+1}\}(\neq l_1)}
\sqrt{\nu ^{(p+1)}(l_{1}, \cdots , l_{p+1})}
\sqrt{\nu ^{(p)}(l_{2}, \cdots , l_{p+1})} 
\Biggr)^{2}.
\label{qend}
\end{eqnarray}
Subtracting $p$ times eq.~(\ref{probend}) from eq.~(\ref{nend}),
we get
\begin{equation}
n-p=-\sum_{\{l_{1},\cdots,l_{p-1}\}}
\nu^{(p-1)}(l_{1},l_{2},\cdots,l_{p-1})
+
\sum_{\{l_{1},\cdots,l_{p+1}\}}
\nu^{(p+1)}(l_{1},l_{2},\cdots,l_{p+1}).
\label{last}
\end{equation}
For the Brinkman-Rice transition that $q(l)$, eq.~(\ref{qend}),
to vanish,
the two terms in the right-hand side of eq.~(\ref{last}) 
must vanish  at a time.
This is because
$\nu ^{(p)}$ in eq.~(\ref{qend}) generally 
remain non-zero to accommodate $n$ particles per site,
or to meet eq.~(\ref{nend}).
(Note that $\nu^{(p)}\ge 0$ for any $p$.)
Thus we conclude that
the Brinkman-Rice transition occurs only
when the carrier density $n$ equals an integer $p$.~\cite{rf:Lu}\


\begin{thebibliography}{99}

\bibitem{rf:Okabef} T. Okabe:
J. Phys. Soc. Jpn. {\bf 65} (1996) 1056.
\bibitem{rf:comment1}
Recently, B\"unemann and Weber came to the same conclusion;
J. B\"unemann and W. Weber: preprint (SISSA: cond-mat/9611031).
\bibitem{rf:Vollhardt}D. Vollhardt:
Rev. Mod. Phys. {\bf 56} (1984) 99.
\bibitem{rf:BR}W. F. Brinkman and T. M. Rice:
Phys. Rev. {\bf B2} (1970) 4302.
\bibitem{rf:Lu}J. P. Lu:
Phys. Rev. {\bf B49} (1994) 5687; preprint (SISSA: cond-mat/9601133).

\end{thebibliography}
\end{document}